\documentclass{aa}  
\usepackage{graphicx}
\usepackage{txfonts}
\usepackage{natbib}
\bibpunct{(}{)}{;}{a}{}{,}
\usepackage{epsfig}
\begin{document}
   \title{The optical counterparts of Accreting Millisecond X-Ray Pulsars during quiescence\thanks{Based on observations made with 
   ESO Telescopes at the Paranal Observatory under programmes ID 077.D-0677(D) and ID 079.D-0884(A) and observations made with the 
   ESO telescopes obtained from the ESO/ST-ECF Science Archive Facility.}}


   \author{P. D'Avanzo
          \inst{1, 2}
	  \and
          S. Campana\inst{1}
          \and
          J. Casares\inst{3}
	  \and
          S. Covino\inst{1}
          \and
          G. L. Israel\inst{4}
          \and
          L. Stella\inst{4}
          }

   \offprints{P. D'Avanzo}

   \institute{INAF, Osservatorio Astronomico di Brera, via E. Bianchi 46, I-23807 Merate (Lc), Italy\\
              \email{paolo.davanzo@brera.inaf.it}
         \and
             Universit\`a degli Studi dell'Insubria, Dipartimento di Fisica e Matematica, via Valleggio 11, I-22100 Como, Italy
         \and
             Instituto de Astrof\'isica de Canarias, 38200 La Laguna, Tenerife, Canary Islands, Spain
         \and
             INAF, Osservatorio Astronomico di Roma, via Frascati 33, I-00040 Monte Porzio Catone, Roma, Italy
             }

   \date{Received; accepted}

 
  \abstract
   {Eight Accreting Millisecond X--ray Pulsars (AMXPs) are known to date. 
   Although these systems are well studied at high energies, very little information is 
   available for their optical/NIR counterparts. Up to now, only two of them, SAX J1808.4$-$3658 
   and IGR J00291+5934, have a secure multi-band detection of their optical counterparts in quiescence.}
   {All these systems are transient Low-Mass X--ray Binaries. Optical and NIR 
   observations carried out during quiescence give a unique opportunity to constrain the nature of the 
   donor star and to investigate the origin of the observed quiescent luminosity at long wavelengths. 
   In addition, optical observations can be fundamental as they ultimately allow us to estimate the compact 
   object mass through mass function measurements.}
   {Using data obtained with the ESO-Very Large Telescope, we performed a deep 
   optical and NIR photometric study of the fields of XTE J1814$-$338 and of the ultracompact systems XTE J0929$-$314 
   and XTE J1807$-$294 during quiescence in order to look for the presence of a variable counterpart. If suitable 
   candidates were found, we also carried out optical spectroscopy.}
   {We present here the first multi-band ($VR$) detection of the optical counterpart of XTE J1814$-$338 in quiescence together 
   with its optical spectrum. The optical light curve shows variability in both bands consistent with a sinusoidal modulation 
   at the known 4.3 hr orbital period and presents a puzzling decrease of the $V-$band flux around superior conjunction that may be interpreted as a partial
   eclipse. The marginal detection of the very faint counterpart of XTE J0929$-$314 and deep upper 
   limits for the optical/NIR counterpart of XTE J1807$-$294 are also reported. We also briefly discuss the results reported in the
   literature for the optical/NIR counterpart of XTE J1751$-$305.}
   {Our findings are consistent with AMXPs being systems containing an old, weakly magnetized neutron star, reactivated as a millisecond radio 
   pulsar during quiescence which irradiates the low-mass companion star. The absence of type I X--ray bursts and of hydrogen and helium lines 
   in outburst spectra of ultracompact ($P_{\rm orb} <1$ hr) AMXPs suggests that the companion stars are likely evolved dwarf stars.}


   \maketitle
%

\section{Introduction}

Some Low Mass X--ray Binaries (LMXBs) exhibit sporadic outburst activity while for most of their time they remain in 
a state of low-level activity; these systems are commonly referred to as X-ray transients 
\citep{White84}. Historically, the classification of X-ray transients is made on the basis 
of their spectral hardness. The outbursts of Soft X-ray Transients (SXTs), characterized by 
equivalent bremsstrahlung temperatures $\leq 15$ keV, are often accompanied by a pronounced 
increase in the luminosity of their (faint) optical counterparts and by the onset of type I 
X-ray burst activity \citep{Maraschi77,Woosley76}. During quiescence, these transient systems are very faint in X-rays 
($10^{32} - 10^{33}$ erg s$^{-1}$) and their optical luminosity drops by as much
as 6-7 mag, giving a unique opportunity for the study of their companion stars. These properties 
clearly associate SXTs with LMXBs containing an old, weakly magnetized, neutron star, 
(for a review see, e.g. \citealt{Campana98a}). In the following we will refer to these systems 
as Low-Mass X--ray Transients (LMXTs).

It has long been suspected that millisecond radio pulsars are the 
spun-up products of sustained mass transfer onto neutron stars in LMXBs. 
According to this scenario, the companions in the LMXBs evolve and transfer 
matter onto the neutron star on a long time-scale, spinning it up to periods 
as short as a few ms. This model has gained strong support about ten years 
ago thanks to the launch of the {\it Rossi X-ray Timing Explorer} ({\it RXTE}) 
satellite that discovered kilohertz quasi-periodic oscillations (kHz QPOs) as
well as coherent oscillations during type-I X-ray bursts in a number of LMXBs 
(e.g. \citealt{Strohmayer96}). These periodicities are interpreted as the millisecond spin periods of 
weakly magnetic neutron stars. In April 1998 the first Accreting 
Millisecond X-ray Pulsar (AMXP) was discovered, with a 401 Hz (2.5 ms) X-ray pulsations 
(SAX J1808.4-3658; \citealt{Wijnands98}). This conclusion was further strengthened 
by the discovery of additional systems in the following years. Eight such 
systems are now known: SAX J1808.4$-$3658 \citep{Wijnands98, Chakrabarty98b}; 
XTE J1751$-$305 \citep{Markwardt02}; XTE J0929$-$314 \citep{Galloway02}; XTE J1807$-$294 
\citep{Markwardt03a}; XTE J1814$-$338 \citep{Markwardt03b, Strohmayer03}; IGR J00291+5934 
\citep{Galloway05}; HETE J1900.1-2455 \citep{Morgan05} and SWIFT J1756.9-2508 \citep{Krimm07}\footnote{Very recently, 
episodes of coherent ms X-ray pulsations were discovered in archival data of the
LMXTs Aql X-1 \citep{Casella08} and SAX J1748.9-2021 \citep{Altamirano08}. We 
will not consider these sources in the present paper.}. 
These findings directly confirmed evolutionary models that link the neutron stars of LMXBs 
to those of millisecond radio pulsars, the former being the progenitors of the latter.
All these systems are LMXTs, have orbital periods in the range 
between 40 min and 4.3 hr and spin frequencies from 1.7 to 5.4 ms. These eight accreting 
millisecond pulsars are well studied at high energies, especially in X$-$rays, both during 
outburst and quiescence (see \citealt{Wijnands05b} for a review). 
On the other hand, with the significant exception of SAX J1808.4$-$3658 and IGR J00291+5934, 
their optical/NIR quiescent counterparts are only poorly known. 
The optical light curve of SAX J1808.4$-$3658 in outburst and quiescence shows variability modulated at
the orbital period, in antiphase with the X$-$ray light curve \citep{Giles99, Homer01, Campana04b}. 
This is unlike other quiescent transients that normally show a double-humped morphology, due to an 
ellipsoidal modulation, and indicates that the companion star is exposed to some irradiation. The same 
behaviour is seen in the quiescent optical light curve of IGR J00291+5934 (D'Avanzo et al. 2007)\footnote{A recent paper of \citet{Jonker08} pointed 
out evidence for a change in the quiescent level of IGR J00291+5934 and for strong flaring episodes. A slight phase difference in the folded 
light curve is also found with respect to that reported in D'Avanzo et al. (2007)}.
In a theoretical study on SAX J1808.4$-$3658 \citet{Burderi03} proposed, on the base of previous works 
(\citealt{Stella94}; Campana et al. 1998), that the irradiation is due to 
the release of rotational energy by the fast spinning neutron star, switched on, as a radio pulsar, during 
quiescence. Following this idea, Campana et al. (2004) and D'Avanzo et al. (2007) measured the required irradiating 
luminosity needed to match the optical flux of SAX J1808.4$-$3658 and IGR J00291+5934, respectively, and found that 
it is a factor of about 100 larger than the observed quiescent X$-$ray luminosity for both systems. Neither accretion$-$driven 
X$-$rays nor the intrinsic luminosity of the secondary star or the disk can account for it. So, these authors conclude 
that the only source of energy available within these systems is the rotational energy of the neutron star, reactivated 
as a millisecond radio pulsar.

Among the eight accretion-powered millisecond X--ray pulsars known to date, four of them are in ultracompact systems, with
orbital periods shorter than 60 minutes. Three of these ultracompact systems, XTE J1751$-$305, XTE J0929$-$314 and XTE J1807$-$294 
are remarkably homogeneous, with measured orbital periods of 42.4, 43.6 and 40.1 minutes respectively, 
well below the minimum period for a system with a donor composed primarily of hydrogen ($P_{\rm orb} \leq 80$ min; \citealt{Rappaport82}). 

Optical and NIR observations performed in the past by different groups only led to deep 
upper limits for the counterparts of XTE J1751$-$305 \citep{Jonker03} and XTE J1814$-$338 \citep{Krauss05} or to the detection of 
very faint candidates, if any (\citealt{Monelli05} for XTE J0929$-$314). No observations of XTE J1807$-$294 during quiescence have been
reported to date, while HETE J1900.1-2455 is still in outburst and SWIFT J1756.9-2508 has been discovered very recently (June 2007). 
The intrinsic faintness of the targets during quiescence, in combination with the large interstellar absorption and high stellar 
crowding of the relevant fields clearly explain the lack of detections at optical wavelengths.


\section{Observations and data reduction}

Optical and NIR observations of the field of XTE J1814$-$338, XTE J0929$-$314 and XTE J1807$-$294 
were carried out with the ESO Very large Telescope (VLT), using FORS1, FORS2 and ISAAC. All nights 
were clear, with seeing in the range $0.5\arcsec{}-1.0\arcsec{}$. For XTE J1814$-$338 we collected 
from the ESO archive\footnote{http://archive.eso.org/} $VR$ images taken with the FORS2 camera on 
nights 2004 May 20-21. A set of about 8 images of 5-6 min integration were obtained every night which 
cover about $30-40$\% of the 4.3 hr orbital period at each filter. 
From the ESO archive we downloaded $VRI$ images of the field of XTE J0929$-$314 
taken with the FORS1 camera on nights 2003 December 19-20-21. The dataset consists of a set of 6 images of 5 min 
integration that cover more than twice per night the 43.6 min orbital period at each filter.
We observed the field of XTE J1807$-$294 with our approved program ID 077.D-0677(D) in service mode with the ISAAC camera. 
Observations were carried out in $J-$band on 2006 May 5 and covered about 92\% of the 40.1 min orbital period of 
the system. In addition, we collected from the ESO archive $VRI$ images taken with the FORS2 camera on night 2004 June 
10 that cover about $30\%-60\%$ of the orbital period. 
We also obtained optical spectra of the counterpart candidates of XTE J1814$-$338 and XTE J1807$-$294 
with the FORS1 camera on nights 2007 September 2-3 during our observing run approved under program ID 079.D-0884(A).
The complete observing log is presented in Table~\ref{tab:log} 

Image reduction was carried out following standard procedures: subtraction of an averaged bias frame,
division by a normalized flat frame; NIR frames were reduced using the ISAAC pipeline data reduction {\it jitter}, 
part of the ECLIPSE\footnote{http://www.eso.org/projects/aot/eclipse/} package. Astrometry was performed using 
the USNOB1.0\footnote{http://www.nofs.navy.mil/data/fchpix/} and the 2MASS\footnote{http://www.ipac.caltech.edu/2mass/} 
catalogues. PSF-photometry was made with the ESO-MIDAS\footnote{http://www.eso.org/projects/esomidas/} daophot 
task for all the objects in the field. The photometric calibration was done against Landolt standard stars 
for $VRI$ filters and against the 2MASS catalog for NIR filters. In order to minimize any systematic effect, 
we performed differential photometry with respect to a selection of local isolated and non-saturated 
reference stars. 

Our spectra of XTE J1814$-$338 and XTE J1807$-$294 were acquired with the grism 300V, covering the
wavelength range 4000$-$9000 \AA . We used a 
$1\arcsec{}$ slit, resulting in an effective resolution of $R = 440$.  The extraction of the spectrum was performed with the 
ESO-MIDAS software package. Wavelength and flux calibration of the spectra were achieved using helium-argon lamp and observing 
spectrophotometric stars.

A short X$-$ray observation of XTE J1814$-$338 was carried out with the XMM-Newton satellite. Details about the observation 
and data analysis are given in Sec. 3.1.

\begin{table*}
   \centering
\caption{Observation log.}
\begin{tabular}{cccccc}
\hline
Source          &  UT observation &  Exposure                   &  Seeing  &   Instrument  & Filter/\\
                &    (YYYmmdd)    &  (s)                        & (arcsec) & 	           & Grism  \\ 
\hline
XTE J1814$-$338 &  20040521.27086 &  $10 \times 60$ s 	        & $0.8''$  & VLT/FORS2	   & $V$    \\
                &  20040521.27194 &  $9 \times 60$ s 	        & $0.8''$  & VLT/FORS2	   & $R$    \\
                &  20040522.26303 &  $10 \times 60$ s 	        & $0.6''$  & VLT/FORS2	   & $V$    \\
                &  20040522.25830 &  $9 \times 60$ s 	        & $0.6''$  & VLT/FORS2	   & $R$    \\
\hline
                &  20070904.07468 &  $4 \times 1800$ s 	        & $0.5''$  & VLT/FORS1	   & $300V$ \\
\hline
\hline
XTE J0929$-$314 &  20031220.27210 &  $6 \times 300$ s           & $0.6''$  & VLT/FORS1     & $V$    \\
                &  20031220.27646 &  $6 \times 300$ s           & $0.6''$  & VLT/FORS1     & $R$    \\
                &  20031220.28083 &  $6 \times 300$ s           & $0.6''$  & VLT/FORS1	   & $I$    \\
                &  20031221.29201 &  $6 \times 300$ s           & $0.7''$  & VLT/FORS1     & $V$    \\
                &  20031221.29636 &  $6 \times 300$ s           & $0.7''$  & VLT/FORS1	   & $R$    \\
                &  20031221.30071 &  $6 \times 300$ s           & $0.7''$  & VLT/FORS1	   & $I$    \\
                &  20031222.26029 &  $6 \times 300$ s           & $0.7''$  & VLT/FORS1     & $V$    \\
                &  20031222.26467 &  $6 \times 300$ s           & $0.7''$  & VLT/FORS1	   & $R$    \\
                &  20031222.26904 &  $6 \times 300$ s           & $0.7''$  & VLT/FORS1	   & $I$    \\
\hline
\hline
XTE J1807$-$294 &  20040610.16380 &  $9 \times 60$ s            & $0.7''$  & VLT/FORS2     & $V$    \\
                &  20040610.17390 &  $9 \times 60$ s            & $0.7''$  & VLT/FORS2     & $R$    \\
                &  20040610.18748 &  $16 \times 60$ s           & $0.7''$  & VLT/FORS2     & $I$    \\
                &  20060505.19875 &  $30 \times 3 \times 20$ s  & $1.0''$  & VLT/ISAAC     & $J$    \\
\hline
                &  20070903.19999 &  $1 \times 600$ s 	        & $0.9''$  & VLT/FORS1	   & $300V$ \\
\hline
\end{tabular}
\label{tab:log}
\end{table*}

%

\section{XTE J1814$-$338}

XTE J1814$-$338 was discovered in outburst on 2003 June 5 during monitoring of the central Galactic plane with 
the {\it RXTE} \citep{Markwardt03b}. The detection of coherent pulsations 
at frequency of 314.3 Hz (3.2 ms), modulated on an orbital period of 4.3 hr, makes this source the widest 
accreting millisecond X--ray pulsar known to date \citep{Markwardt03b, Markwardt03c}. Many X$-$ray bursts were 
observed from XTE J1814$-$338 with the {\it RXTE} satellite. Assuming the Eddington limit for the brightest of 
those bursts \citet{Strohmayer03} obtained a source distance of $8.0 \pm 1.6$ kpc. X--ray observations carried out
with the {\it Chandra} satellite during outburst led to a precise source position and showed a featureless X-ray
spectrum which is best fit by an absorbed power law plus blackbody model, where the equivalent hydrogen column 
density ($N_H$) is comparable to the Galactic value \citep{Krauss05}. An $R \sim 18.3$ candidate optical 
counterpart was identified within the small {\it Chandra} error box during the outburst and was no longer visible
during quiescence down to a limiting magnitude $R > 23.3$ ($3\sigma$ c.l.; Krauss et al. 2005). Optical 
spectroscopy of the candidate carried out during the outburst revealed prominent hydrogen and helium emission lines, 
supporting its association with XTE J1814$-$338 \citep{Steeghs03}. An X-ray heated accretion disk model is able to 
account for the $BVR$ magnitudes measured during the 2003 outburst but this model is unable to account for the 
$I-$band data, which are systematically brighter than predicted (Krauss et al. 2005). A similar ``IR excess'' 
with respect to the X-ray heated disk model, was observed in the accretion-powered millisecond pulsar 
SAX J1808.4-3658 (Wang et al. 2001; Greenhill et al. 2006) and suggests for both sources a significant contribution 
from the companion star or from a synchrotron-emitting region \citep{Wang01, Krauss05, Greenhill06, Russell07}. 
No radio observations of this source have been reported to date. 
Using {\it RXTE} data taken during outburst, \citet{Papitto07} performed a precise timing analysis measuring 
a spin-down behaviour of the source ($\dot{\nu} \sim -7 \times 10^{-14}$ Hz s$^{-1}$) and used this value to 
estimate a magnetic field of $10^8 - 10^9$ G. No quiescent X--ray observations of this system have been reported to date.

\subsection{XMM$-$Newton observations}

XMM-Newton observed XTE 1814--338 on Sep. 6 2005 for 41 ks. 
The observation is plagued by strong proton flares and only 
9 ks (pn data) survive after filtering. A faint source can be 
detected at the known source position with a signal to noise ratio 
of 2.6. The source count rate is $(4.9\pm1.9)\times 10^{-3}$ c s$^{-1}$ 
(1 $\sigma$ confidence level).
The source is not detected in the MOS1 and MOS2 exposures.
Assuming a power law spectrum with $\Gamma=2$ and the outburst 
column density (Krauss et al. 2005) we can estimate a 0.5--10 keV
unabsorbed flux of $\sim 3\times 10^{-14}$ erg cm$^{-2}$ s$^{-1}$,
converting to a quiescent luminosity of $\sim 2\times 10^{32}$ erg 
s$^{-1}$ (assuming a distance from Earth of 8 kpc). Despite that this 
luminosity is somewhat higher than what usually observed in AMXPs, we 
remark that the uncertainties connected to this value are very large due 
to the paucity of counts as well as to the spectral parameters.

\subsection{Optical counterpart during quiescence}

\begin{figure}
\epsfig{file=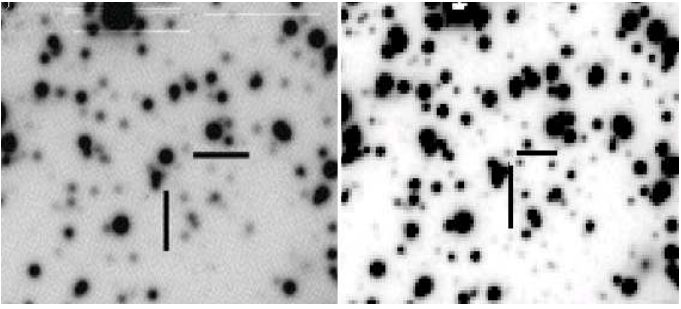, width=9.0cm, height=4.0cm, angle=0}
\vskip 0.0truecm
      \caption{Finding chart for XTE J1814$-$338 ($30'' \times 30''$). The detection of the source during outburst (left
panel, from Krauss et al. 2005) and $R-$band VLT image taken during quiescence (right panel). The source is
dimmer but still clearly visible.
              }
         \label{fig:fc}
   \end{figure}

In all our $V$ and $R$ frames we found a point-like object inside the {\it Chandra} error box at the 
following coordinates (J2000): R.A. = $18^h 13^m 39^s.04$, Dec = $-33^{\circ} 46' 22''.3$ ($0.3''$ error). 
This position is coincident with the one of the optical counterpart of XTE J1814$-$338 detected by Krauss 
et al. (2005) during the 2003 outburst. The object has $V \sim 23.3$ and $R \sim 22.5$. 
A finding chart is shown in Fig.~\ref{fig:fc}.

\subsection{Optical light curve}

\begin{figure}
 \includegraphics[width=\columnwidth]{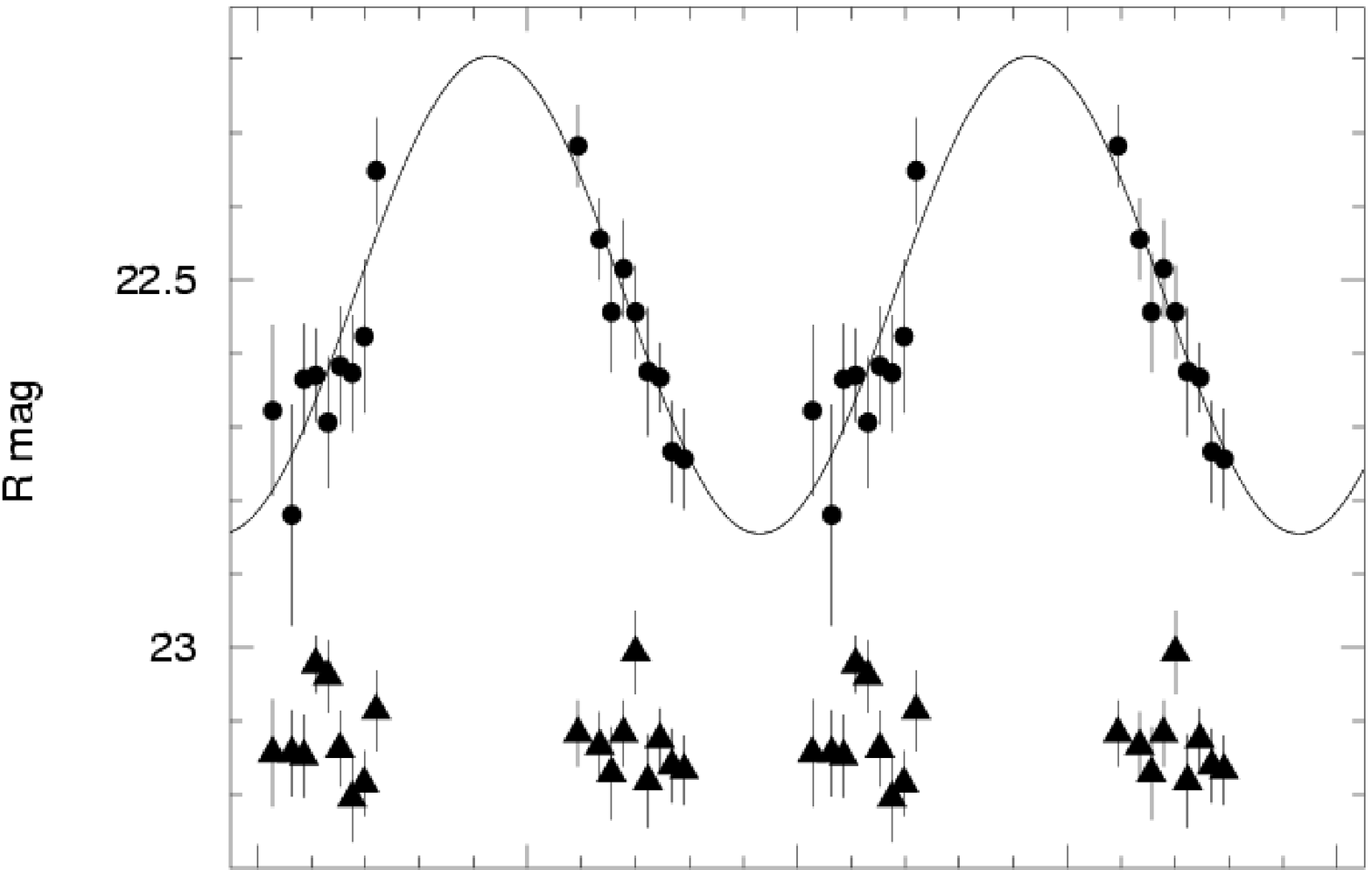}\\%
 \includegraphics[width=\columnwidth]{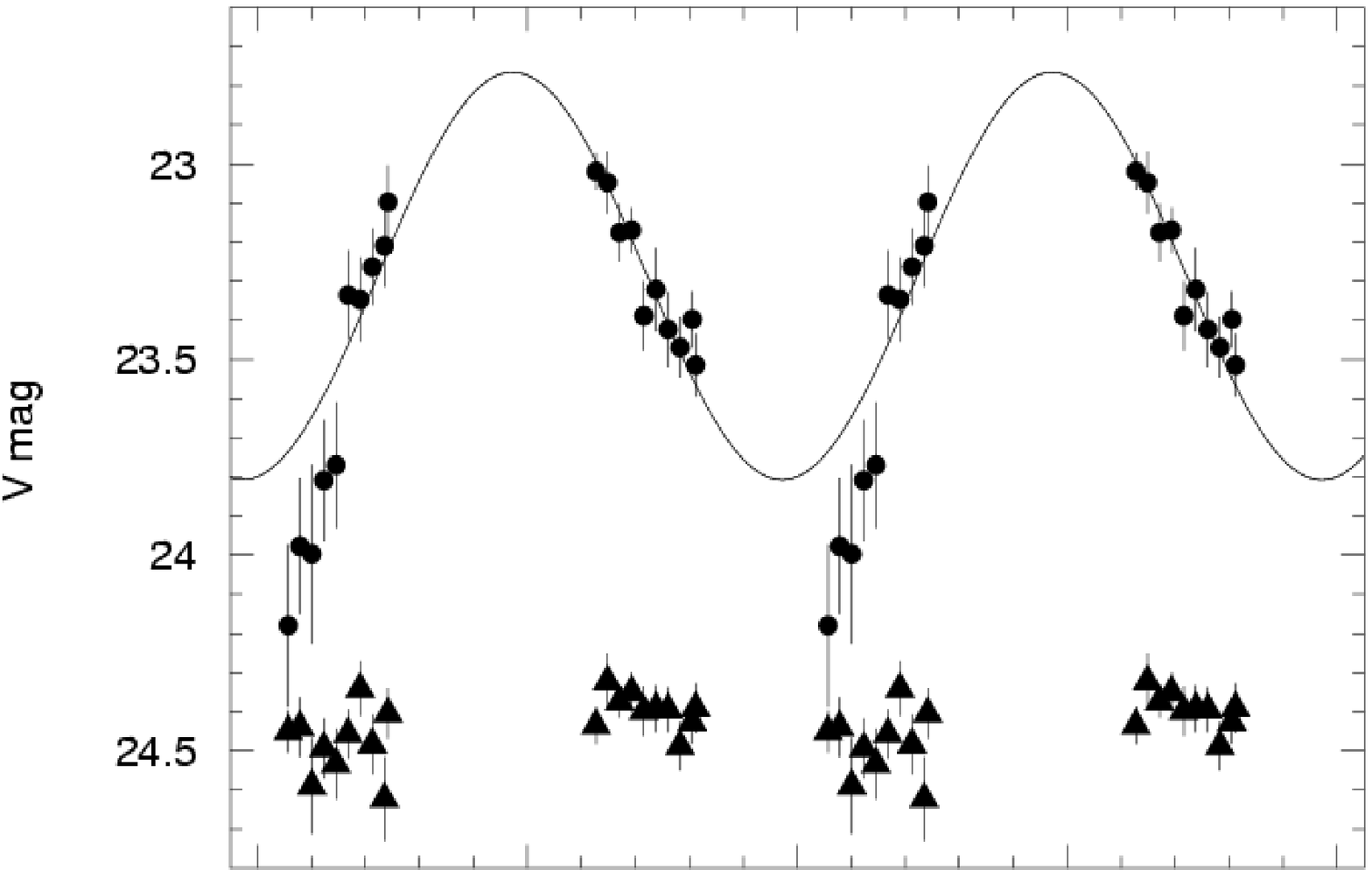}
 \includegraphics[width=\columnwidth]{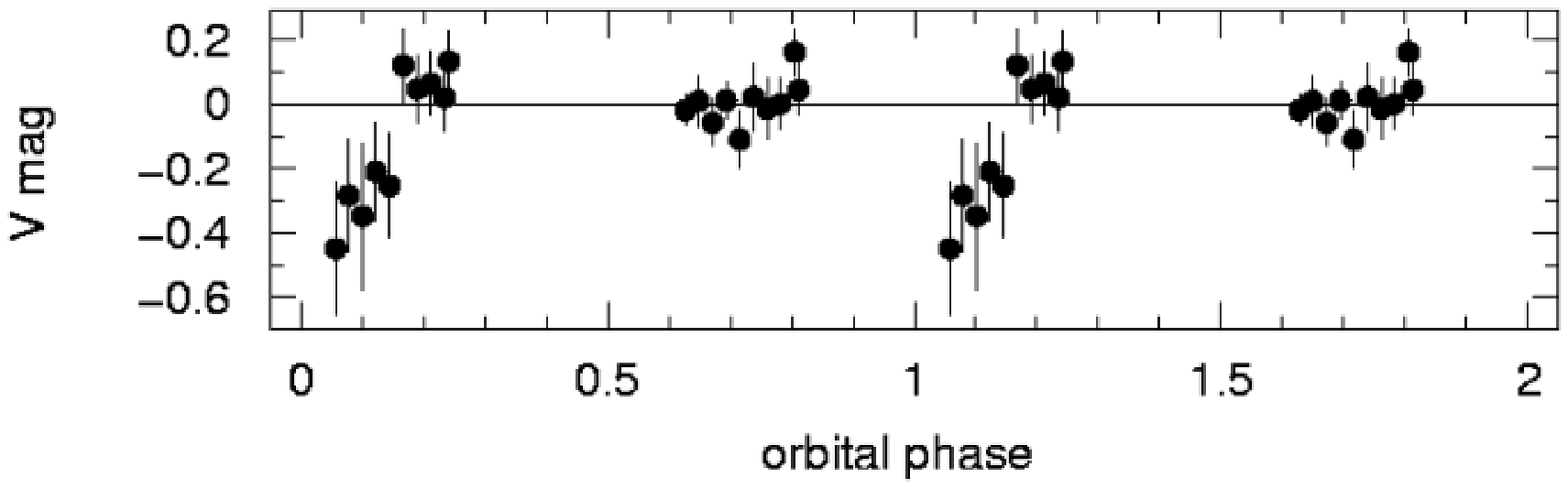}
  \caption{$R$ and $V-$band light curves of XTE J1814$-$338. Phase zero is defined as superior conjunction 
      (i.e. when the neutron star is behind the companion) and is computed on the X--ray ephemerides of 
      Papitto et al. (2007). Two orbital phases are shown for clarity, 
      the best sine$-$wave fit is also shown. The light curves for a 
      field star of comparable brightness are plotted offset below (triangles). The residuals of the 
      $V-$band light curve to the sinusoid model are shown in the lower panel.}
         \label{fig:lc}
\end{figure}

Once we identified the candidate, we searched for variability of this source in each filter. We performed PSF-photometry 
of our candidate and of a selected sample of bright, non-saturated, isolated stars assumed to be non-variable 
in all our $VR$ frames. The result of such differential photometry is
that our source is variable with a clear sinusoidal modulation at the 4.3 hr orbital period with an average
semiamplitude of about 0.4 mag (Fig.~\ref{fig:lc}). This unambiguously identifies the source as the optical 
counterpart of XTE J1814$-$338 and represents the first detection at optical wavelengths of this source during 
quiescence. In both $V$ and $R-$band the light curve shows a single minimum around phase 0, i.e. at superior conjunction 
(when the neutron star is behind the companion) and a maximum around phase 0.5 ($0.48 \pm 0.02$ and $0.43 \pm 0.02$ 
respectively, based on the precise X--ray ephemerides of Papitto et al. 2007), likely indicating that irradiation 
of the companion star plays a crucial role and it is reminiscent of the optical light curves of SAX J1808.4$-$3658 and 
IGR J00291+5934 during quiescence. All the results of our phase-resolved multi-band photometry are reported in 
Table~\ref{tab:phot}.

It is interesting to note that in the $V-$band light curve, the points between phases 0.05 - 0.17 are well below 
the sinusoid best-fit of the data. We carefully checked the $V-$band images used to obtain the measurement, and it seems that 
they are not affected by problems (e.g., cosmic rays, spikes, bad pixels). In addition, the comparison stars are 
not affected by this decrease.  
However, there is no clear evidence of a similar effect in the corresponding points of the $R-$band light curve taken 
contemporaneously (Fig.~\ref{fig:lc}).
A tentative explanation for this observed phenomenon will be presented and discussed in Sec. 3.5.

\begin{table}
\caption{Results of photometry of XTE J1814$-$338, all the values are uncorrected for reddening. In column five are reported the 
reddening parameters used to correct our optical photometry, computed assuming $E(B-V)=0.29 \pm 0.03$ mag 
(see Sec. 3.3 for details).}
\begin{tabular}{ccccc}
\hline
Filter &${\lambda}_c$&  Mean magnitude   &   Semiamplitude  &  A$_{\lambda}$  \\ 
       &   (\AA)     &  		 &	 (mag)      &	  (mag)       \\ 
\hline
$V$    &  5270       & $23.29 \pm 0.04$  & $0.52 \pm 0.08$  & $0.87 \pm 0.03$ \\
$R$    &  6440       & $22.52 \pm 0.03$  & $0.32 \pm 0.08$  & $0.68 \pm 0.03$ \\
\hline
\end{tabular}
\label{tab:phot}
\end{table}

\subsection{Optical spectrum}

\begin{figure}
\epsfig{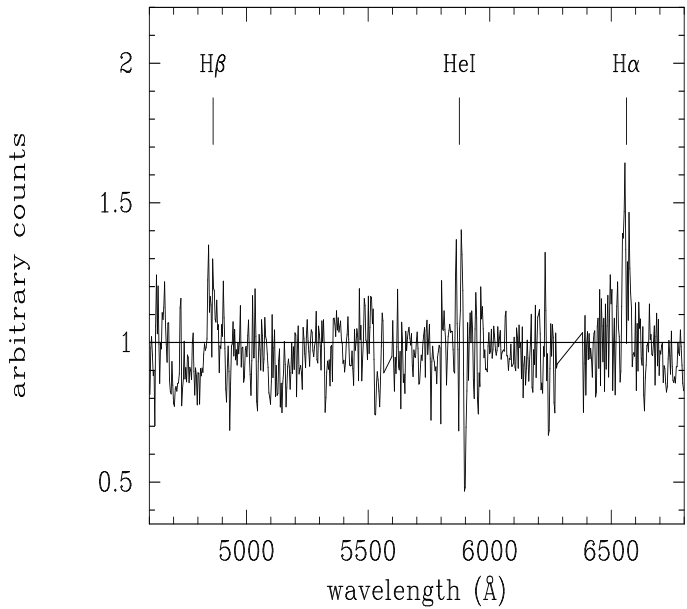}
\vskip 0.0truecm
      \caption{Average optical spectrum of XTE J1814$-$338 during quiescence. We masked some bad-subtracted sky emission lines. 
      The flux is normalized to continuum. H$\alpha$, H$\beta$ and a possible HeI $\lambda$5875 emission lines are visible.
                    }
         \label{fig:spe}
   \end{figure}

Our spectra of XTE J1814$-$338 were acquired on the night 2007 Sept. 3, with the FORS1 camera of the 
ESO-VLT equipped with the grism 300V, covering the wavelength range 4000$-$9000 \AA.
We calibrated them observing spectrophotometric standard stars and corrected the flux for interstellar absorption 
(see Sec. 3.5 for details on reddening correction). We accounted for slit losses by matching our $V$ and $R$ photometry. 
With a simple rescaling (by a factor 1.2) we were able to correct the flux discrepancy between spectra and photometry, 
showing that the spectral flux calibration was robust. We then computed an average of all the available spectra, in order 
to increase the signal-to-noise ratio. In our averaged spectrum we can clearly detect H$\alpha$, H$\beta$ and a 
possible HeI $\lambda$5875 emission lines, superposed on a blue continuum (Fig.~\ref{fig:spe}), with equivalent widths of 
$(19.0 \pm 1.1)$ \AA , $(36.4 \pm 3.6)$ \AA \,\,and $(9.4 \pm 1.2)$ \AA , respectively. These emission lines are also 
visible in each single spectrum. Unfortunately, the S/N and the spectral resolution are not high enough to determine if 
these lines have a double-horned profile (the typical signature of an accretion disk) or not.

\subsection{Origin of the quiescent optical emission}

\begin{figure}
\epsfig{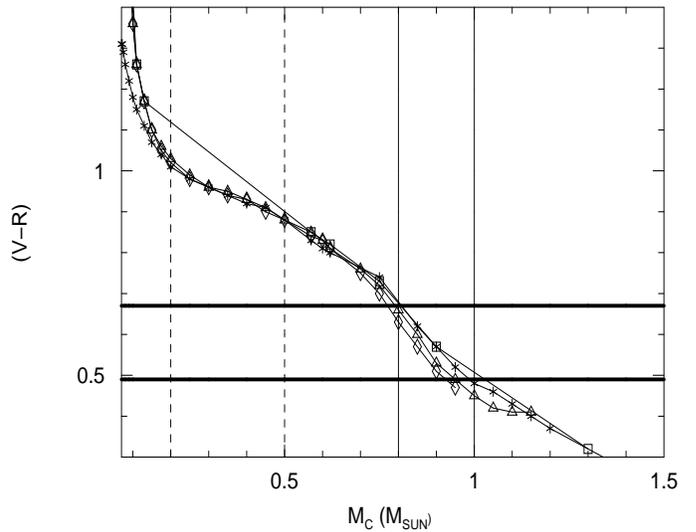}
\vskip 0.0truecm
     \caption{Mass-color diagram for the optical counterpart of XTE J1814$-$338. Isochrones (from Baraffe et al. 1998) 
     are plotted for low-mass stars with ages 0.1 (stars), 0.5 (squares), 5 (triangles) and 10 (diamonds) Gyr. 
     Horizontal and vertical solid lines delineates our measure of the color and the derived estimate of the companion mass, respectively. 
     Vertical dashed lines mark the mass estimates obtained by Krauss et al. (2005).
	     }
	\label{fig:CM}
  \end{figure}

The X--ray mass function ($f_X(M) \sim 2.0 \times 10^{-3} M_{\odot}$) computed by \citet{Markwardt03c}, combined with a neutron star mass of 
$1.4 \, (2.0) \, M_{\odot}$, implies for XTE J1814$-$338 a minimum companion mass of $0.17 \, (0.21) \, M_{\odot}$.  
Based on the harmonic properties of the burst oscillations, \citet{Bhattacharyya05} 
were able to derive 90\% confidence intervals of $26^{\circ} < i < 50^{\circ}$ for the 
inclination of XTE J1814$-$338. A similar constraint was derived by Krauss et al. (2005) from 
the absence of X--ray eclipses or dips and assuming an absolute magnitude of the system during 
quiescence of $M_V < 13.2$. With these constraints on the system inclination the companion 
mass should be in the range $(0.2 \leq M_2 \leq 0.5) \, M_{\odot}$. 
Information about the companion mass can also be obtained from our color photometry. 
To compute the color of the source we first need to correct our magnitudes for interstellar absorption 
computed with the relation $N_H/E(B-V) = 5.8 \times 10^{21}$ cm$^{-2}$ mag$^{-1}$ \citep{Bohlin78} and 
assuming $N_H = (1.67 \pm 0.17) \times 10^{21}$ cm$^{-2}$ (Krauss et al. 2005). The resulting color excess is 
$E(B-V)=0.29 \pm 0.03$ mag. Using a standard extinction curve from \citet{Fitzpatrick99} we obtained the reddening 
parameters for our filters (Table~\ref{tab:phot}).
We thus computed a mean unabsorbed $(V-R) = 0.58 \pm 0.09$ color and compared it (Fig.~\ref{fig:CM}) with the theoretical mass-color 
diagrams computed for solar metallicity low-mass stars reported in \citet{Baraffe98}. 
The result is that the secondary's mass should be $0.8 \leq M_c \leq 1.0$ $M_{\odot}$ ($1\sigma$ c.l.). 
Using the X$-$ray mass function, we can derive for these masses the inclination of the system 
$13^{\circ} \leq i \leq 15^{\circ}$. Assuming a uniform a priori distribution of inclination angles, the inferred inclination 
would be low, and therefore improbable ($P \sim 3\%$). In addition, these inclinations are at variance with the estimates 
of Bhattacharyya et al. (2005) and Krauss et al. (2005) described above.

\begin{figure}
\epsfig{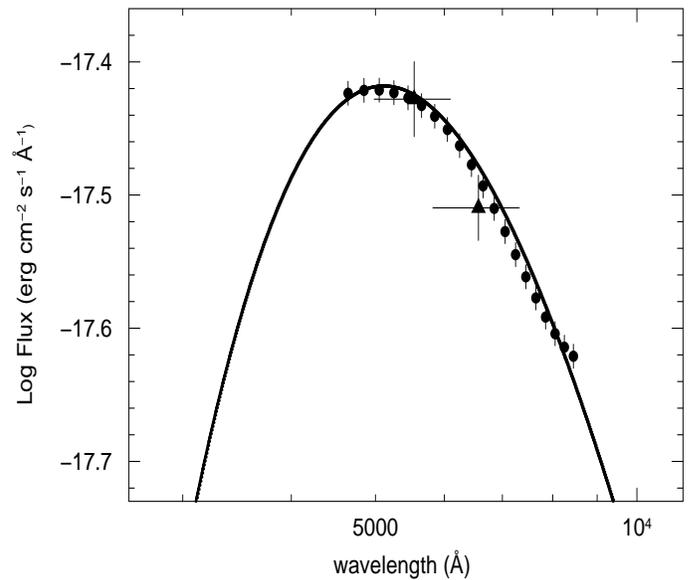}
\vskip 0.0truecm
      \caption{Optical spectrum (dots) and $VR$ broadband photometry (triangles) of XTE J1814$-$338 during quiescence. 
      Data are corrected for interstellar absorption. The plotted line represents the contribution of an irradiated star. 
      See Sec. 3.5 for details.
              }
         \label{fig:sed2}
   \end{figure}

The discrepancies in the estimate of the companion's mass can find a natural explanation under the hypothesis that the
secondary star is subjected to strong irradiation from the compact object, as suggested by the optical light curve. 
Since during our observations XTE J1814$-$338 was in quiescence, we try to investigate the possible causes of its optical/NIR 
emission. Following Campana et al. (2004) and \citet{Davanzo07} we now attempt to account for the optical spectrum flux
(corrected for interstellar absorption) with the simple model of an irradiated star with a blackbody 
spectrum (for the details of the modeling see \citealt{Chakrabarty98a} eqs. [8]$-$[13]). We fit the data by using the irradiating 
luminosity ($L_{irr}$), the distance from Earth ($D$), the radius of the companion star ($R_c$) and the albedo of the star (${\eta}_*$) 
as free parameters. To fill the Roche lobe of a 4.3 hr binary system, a main-sequence companion star of XTE J1814$-$338 
should obey the mass-radius relation $R_c = 0.28(M_c/0.1M_{\odot})^{1/3}R_{\odot}$ (see, e.g., \citealt{Frank92}). 
For each possible radius $R_c$ we can thus derive the binary separation $a~=~(G(M_X~+~M_c)(P_{\rm orb})^2/(4\pi)^2)^{1/3}$ 
which is also a parameter foreseen in the Chakrabarty (1998) model. We obtain an acceptable fit to all the data 
(see Fig.~\ref{fig:sed2} and Table~\ref{tab:fit}), with a reduced ${\chi}^2 = 0.9$ (18 degrees of freedom, n.h.p. of 58\%). 

\begin{table*}
   \centering
\caption{Fit parameters used to model the XTE J1814$-$338 optical spectrum (see Sec. 3.5 for details). Errors are at 90\% confidence level.}
\begin{tabular}{ccccccc}
\hline
Model        &  $L_{irr}$                        & Distance           &  $R_c$           &  ${\eta}_*$      &  R$_{in}$                    & ${\chi}^2$  \\ 
             &  (erg s$^{-1}$)                   &   (kpc)            & ($R_{\odot}$)    &	            &  (cm)                        &             \\ 
\hline
Star         &  $(1.02 \pm 0.08)\times 10^{34}$  &  $11.75 \pm 0.54$  & $0.37 \pm 0.02$  & $0.16 \pm 0.07$  & $-$                            & 0.9   \\
Disc + star  &  $(2.74 \pm 0.03)\times 10^{34}$  &  $10.25 \pm 0.10$  & $0.34 \pm 0.01$  & $0.75 \pm 0.01$  & $< 1.8 \times 10^{8}$ & $< 0.8$   \\
\hline
\end{tabular}
\label{tab:fit}
\end{table*}

\noindent The result of our fit is that the required irradiating luminosity is $L_{irr} \sim 1 \times 10^{34}$ erg s$^{-1}$. 
Taking this value as a lower limit for the spin-down luminosity of a classical rotating magnetic dipole, we can estimate a 
neutron star's magnetic field greater than $9.7 \times 10^7$ Gauss. Such value is in agreement with the estimate of Papitto et al. 
(2007). 
The required irradiating luminosity is about two orders of magnitude larger than the quiescent X$-$ray 
luminosity of XTE J1814$-$338 ($L_{X} \sim 10^{32}$ erg s$^{-1}$; see Sec. 3.1). Such discrepancy is 
reminiscent of that observed for SAX J1808.4$-$3658 (Burderi et al. 2003; Campana et al. 2004) and IGR J00291+5934 
\citep{Davanzo07} and can be explained with the presence in the system of a relativistic particle wind from an active 
pulsar which irradiates the companion star.
As reported in Table~\ref{tab:fit}, we were able 
to estimate $R_c = (0.37 \pm 0.02) \, R_{\odot}$ and consequently, using the mass-radius relation reported above, 
$M_c = (0.23 \pm 0.08) \, M_{\odot}$. Such values are typical of a main-sequence M-type star, in agreement with the 
estimate made by Krauss et al. (2005) for this system. On the other hand, the 
$(V-R)$ color we measured is consistent with a late G or K-type main-sequence star. In addition, the optical spectrum of 
XTE J1814$-$338 is well fitted with a blackbody with ${\lambda}_{max} = (5106 \pm 24)$ \AA \, (Fig.~\ref{fig:sed2}), 
equivalent to a surface temperature of $T \sim 5675$ K (typical of a G-K star). With a typical mass and radius comparable 
to the Sun, a G or K star cannot fit the Roche lobe of a binary system like XTE J1814$-$338. However, this discrepancy 
between the observed spectral type of the companion star and the geometrical limits of the system finds a natural explanation 
if we assume that the relativistic particle wind of the radio pulsar, which irradiates the companion, would increase its surface 
temperature making it appear as an earlier spectral type star. 

   \begin{figure}
\epsfig{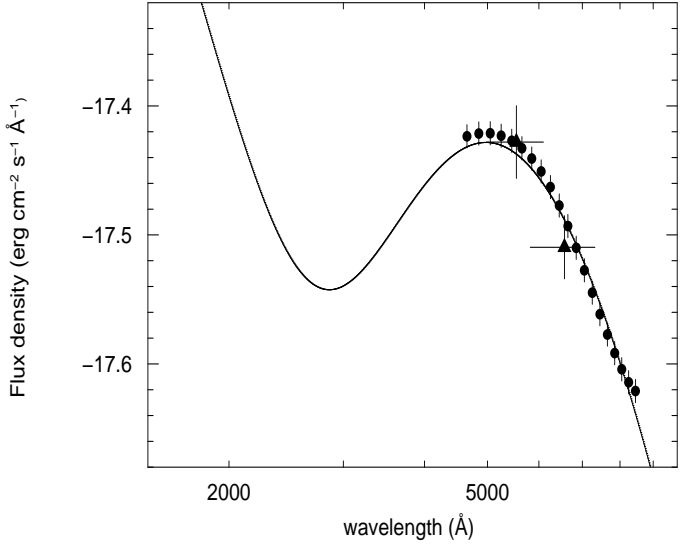}
\vskip 0.0truecm
      \caption{Optical spectra (dots) and $VR$ broadband photometry (triangles) of XTE J1814$-$338 during quiescence. 
      Data are corrected for interstellar absorption. The plotted line represents the contribution of an irradiated star plus a residual accretion disk. 
      See Sec. 3.5 for details.
              }
         \label{fig:sed_bd}
   \end{figure}

The simple model we considered above does not predict the presence of a residual disk during quiescence. 
To check for a more realistic solution we tried to fit our data with the model of an irradiated star plus a disk by using 
the irradiating luminosity ($L_{irr}$), the distance from Earth ($D$), the radius ($R_c$) and the albedo of the companion star 
(${\eta}_*$) and the internal disk radius ($R_{in}$) as free parameters. 
We fixed the X$-$ray albedo of the disk to 0.95 (Chakrabarty 1998). For each possible radius $R_{in}$ we assumed an outer disk radius of $0.3a$ 
(where $a$ is the binary separation). 
We obtained an acceptable solution for values of the fitting parameters similar to the model of the irradiated star alone 
(see Table~\ref{tab:fit} and Fig.~\ref{fig:sed_bd}). We can provide just an upper limit for the inner disk radius, given that the reduced 
${\chi}^2$ of the fit reaches a minimum for values of the inner disk radius that go beyond the level of the neutron star surface. 
Within the limits of our fit, the residual accretion disk contributes about 20\% to the observed optical flux\footnote{We 
also tried to fit our data with a disk model, but we found an acceptable solution only for large
values of the irradiation luminosity ($\sim 10^{36}$ erg s$^{-1}$) and of the disk inner radius ($\sim 10^{10}$ cm).}. Such scenario 
might shed light on the puzzling optical light curve we presented in Sec.~3.2 and in Fig.~\ref{fig:lc}. A possible way 
to explain this $V$ flux decrease can be found assuming that the companion star or an intra-binary shock front (created by the 
interaction of the relativistic pulsar wind and matter outflowing from the companion star and the disk) are eclipsing the disk. 
Since the disk is very blue (peaking at $\lambda < 1000$ \AA \,), we would expect a marked decrease in the $V$ band ($\sim 0.2$ mag) 
and a minor decrease ($\sim 0.1$) in the $R-$band\footnote{Assuming that around phase zero the whole contibution of the disk ($\sim 20\%$) 
to the observed optical flux is eclipsed.}, which might be hidden in the error bars of our photometry. 
As reported in Sec. 3.2, the eclipse duration is of at least 10\% of $P_{\rm orb}$ 
(unfortunately we do not have data between orbital phase $0.80 - 1.05$). We can also derive an upper limit of $0.3P_{\rm orb}$ 
on the eclipse duration assuming that the eclipse profile is symmetric around phase 1.0 (Fig.~\ref{fig:lc}). So, the size of the region 
producing the eclipse should be $(2{\pi}a/P_{\rm orb})0.1P_{\rm orb} \leq R_e \leq (2{\pi}a/P_{\rm orb})0.3P_{\rm orb}$ which means 
$0.6a \la R_e \la 1.9a$, where $a~=~(G(M_X~+~M_c)(P_{\rm orb})^2/(4\pi)^2)^{1/3}$ is the binary separation. Using $M_X = 1.4\,M_{\odot}$ 
and $M_c = 0.2\,M_{\odot}$ we obtain $1.0\,R_{\odot} \la R_e \la 3.0\,R_{\odot}$. This might indicate that 
the extended region is producing the eclipse, rather than a compact feature as also suggested 
by the broad decrease in the observed $V-$band flux. However, if the companion is eclipsing 
the disk, the system should be close to edge-on but, as noted by Krauss et al. (2005), no eclipses or dips 
are visible in the X$-$ray light curve ($i < 77^{\circ}$), weakening the hypothesis of the companion eclipsing the disk.

Alternatively, as shown in Campana et al. (2004) for SAX J1808.4-3658, the intra-binary shock front (if present)
is expected to have a blue spectrum, partially contributing to the whole observed optical flux.
If the companion star is eclipsing this shock front, then the eclipse should be 
visible also at low-inclinations, the shock front being spatially close to the companion star. 
As in the previous scenario, such an eclipse should be more marked in the bluer bands, however only the lower limit of 
our estimate of $R_e$ is compatible with the expected size of the companion star of XTE J1814$-$338.

Finally, if the companion star is not fully convective (but see Podiadlovski 1991)
and has a temperature gradient between the hemisphere facing the 
neutron star and the other one persistently hidden, this can explain the flux decrease without invoking 
eclipses.


\section{Ultracompact Accreting ms X--ray Pulsars}

   \begin{figure*}[!htb]
   \centering
\epsfig{file=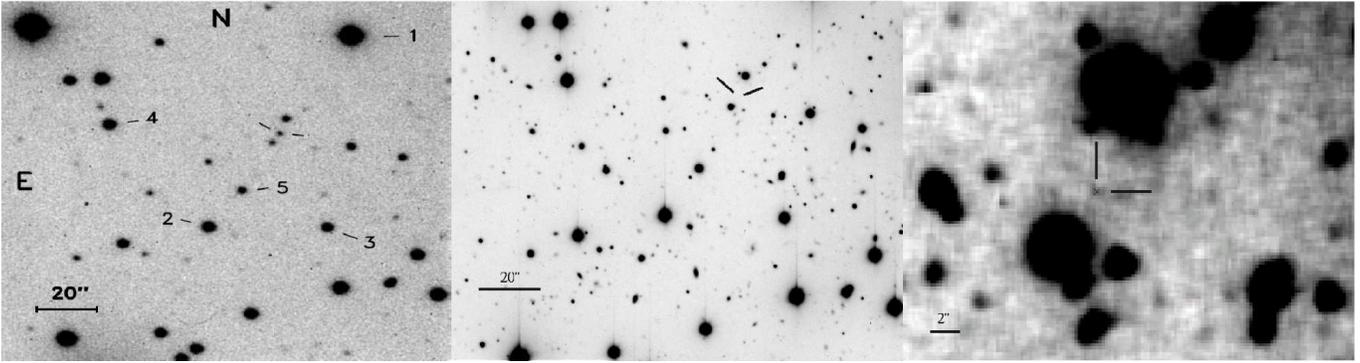, width=18.0cm, height=4.8cm, angle=0}
\vskip 0.0truecm
      \caption{Finding chart of XTE J0929$-$314. The detection of the source during outburst (left
panel, from Giles et al. 2005) and VLT image taken during quiescence (central panel). In the right panel we show a detail of the VLT
image, with the radio position (Rupen, Dhawan \& Mioduszewski 2002) marked by a cross. The optical counterpart of XTE J0929$-$314 
(marked with solid lines) is dimmer but still visible.
              }
         \label{fig:fc0929}
   \end{figure*}

Binary systems with $P_{\rm orb} < 80$ minutes can be formed in at least two possible channels as summarized, e.g., by \citet{Deloye03}. 
The first scenario starts from stable mass transfer onto a neutron star from an evolved main-sequence star close to core hydrogen exhaustion
\citep{Nelson86, Fedorova89, Podsiadlowski02, Nelson03}, or 
a He-burning star \citep{Savonije86, Tutukov89}. According to this scenario, the system goes through 
the 40 minute period range twice. Such a system will initially evolve to orbital periods comparable 
to the ultracompact AMXPs and can reach a period minimum of about 10 minutes. The resulting ultracompact binaries have donor masses of 
$0.1-0.2 \, M_{\odot}$ as they pass through $P_{\rm orb} \sim 40$ minutes on their way toward the minimum period. At this stage, the donors have 
masses significantly greater with respect to those measured for the ultracompact AMXPs known to date. However, systems evolving through 
$P_{\rm orb} \sim 40$ minutes on the way out from their period minimum have masses of the order of 0.01 $M_{\odot}$, more in line with the measurements. 
By this time the donors have become partially degenerate low-mass, low-luminosity donor stars. 
On the contrary, the second channel involves evolution through the common-envelope phase during an unstable mass transfer from an intermediate mass 
($2-6 \, M_{\odot}$) to the neutron star. The donor and the neutron star spiral-in to shorter orbital periods until the envelope is expelled. 
The remaining core of the donor could be either an He or C/O white dwarf star with $M \sim 0.01 M_{\odot}$ \citep{Bildsten02, Deloye03}. The main difference 
between these two evolutionary models is that according to the first one a residual fraction of hydrogen still should be present in the companion star as 
the system reaches $P_{\rm orb} \sim 40$ minutes. Since the evolution of these systems is still unknown, a detection of the companion star will allow us to 
differentiate between these different formation scenarios.

\subsection{XTE J0929$-$314}

 The X$-$ray transient XTE J0929$-$314 was discovered by the All-Sky Monitor (ASM) on the {\it RXTE} satellite in 2002 April \citep{Remillard02}. 
{\it RXTE} observations revealed persistent 185.1 Hz (5.4 ms) pulsations and a binary orbital period of 43.6 minutes. The measured 
X-ray mass function ($f_X(M) = 2.7 \times 10^{-7} M_{\odot}$) gives a minimum companion mass of
$M_c = 0.008 M_{\odot}$ ($i = 90^{\circ}, M_X = 1.4 M_{\odot}$) and implies $M_c \leq 0.03 M_{\odot}$ (95\% c.l.) for a uniform a priori
distribution of inclination angles (Galloway et al. 2002). {\it Chandra} observations carried out during the 2002 outburst showed a
featureless X$-$ray spectrum ($0.5-8.3$ keV) that is well fitted by a power-law $+$ blackbody model with $N_H$ consistent to Galaxy 
line-of-sight values $(0.8-1.0) \times 10^{21}$ cm$^{-2}$ \citep{Juett03}. Variable optical \citep{Greenhill02, Cacella02} and radio 
\citep{Rupen02} counterparts were detected during outburst. Optical photometry performed by 
\citet{Giles05} during outburst revealed a sinusoidal modulation probably due to a combination of emission by a hotspot and an 
X$-$ray heated secondary star. These authors also report the evidence of an excess in the $R$ and $I$ band in the $BVRI$ broad-band 
spectrum (similar to what observed in SAX J1808.4$-$3658 and XTE J1814$-$338), that may be due to synchrotron emission in matter 
flowing out of the system. The optical spectrum of XTE J0929$-$314, carried out during outburst, shows a strong feature around 
4640 \AA , probably due to C or N emission, and no clear evidence for H/He emission or absorption lines \citep{Nelemans06}. 
During quiescence, XTE J0929$-$314 was barely detected by {\it Chandra} in the 0.5$-$10 keV range. The quiescent X$-$ray spectrum could
be fitted with a simple power-law model and no thermal component was detected \citep{Wijnands05a}. Deep optical observations carried
out during quiescence led to the detection of a very faint ($R \sim 27.2$) source at a position consistent with the X$-$ray and radio error
circles (Monelli et al. 2005).

\subsubsection {Optical counterpart during quiescence}

We averaged all our $V$, $R$ and $I$ frames in order to obtain a deep image of the field of XTE J0929$-$314 for each filter. 
We found a faint, point-like object inside the precise radio error circle \citep{Rupen02} at the 
following coordinates (J2000): R.A. = $09^h 29^m 20^s.19$, Dec = $-31^{\circ} 23' 03''.2$ ($0.3''$ error). This position is 
coincident with the one of the optical counterpart detected by Giles et al. (2005) during the 2002 outburst (Fig.~\ref{fig:fc0929}). 
The source is not visible in the single VLT frames, and no check for variability could be performed. 
The results of our PSF $VRI$ photometry are shown in Table~\ref{tab:photJ0929}; we note that we only have a marginal 
detection of the source but our photometry is in agreement with the results reported by Monelli et al. (2005).

\begin{table}
\caption{Results of PSF-photometry of XTE J0929$-$314, values in column three are uncorrected for reddening. In column four we list the 
reddening parameters used to correct our optical photometry computed assuming $E(B-V)=0.15 \pm 0.02$ mag. The absolute magnitudes are 
computed assuming a distance of 8 kpc (Wijnands et al. 2005; see Sec. 4.1.1 for details).}
\begin{tabular}{ccccc}
\hline
Filter &${\lambda}_c$&  Mean magnitude &  A$_{\lambda}$  &   Absolute mag   \\ 
       &   (\AA)     &  	       &     (mag)	 &	            \\ 
\hline
$V$    &  5270       & $28.2 \pm 0.4$& $0.46 \pm 0.02$   & $13.2 \pm 0.4$   \\
$R$    &  6440       & $27.1 \pm 0.3$& $0.36 \pm 0.02$   & $12.3 \pm 0.3$   \\
$I$    &  7980       & $26.9 \pm 0.4$& $0.28 \pm 0.02$   & $12.1 \pm 0.4$   \\
\hline
\end{tabular}
\label{tab:photJ0929}
\end{table}

We can use our detection to evaluate the possible nature of the optical counterpart of XTE J0929$-$314 in light of the possible 
evolutionary states of ultracompact systems presented in Sec. 4. Given that the source is detected
during quiescence, we will assume that, if present, the accretion disk contribution is low with respect to the total observed flux, 
as discussed for XTE J1814$-$338 (Sec. 3.5) and found for other accretion powered millisecond X--ray pulsars (Campana et al. 2004; 
D'Avanzo et al. 2007). 

   \begin{figure}
\epsfig{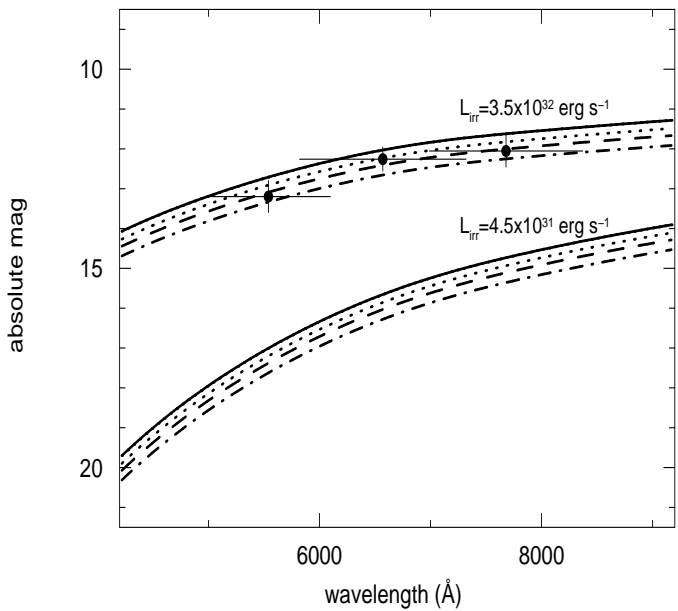}
\vskip 0.0truecm
      \caption{Absolute $VRI$ magnitudes of the optical counterpart of XTE J0929$-$314 (dots) compared to the theoretical models for irradiated
      donor stars in $\sim$ 40-min period ultracompact X--ray binaries (computed for different values of the irradiating luminosity. See Sec.
      4.1.1 for details). Solid lines: the post-period-minimum model for XB 1832-330 (Podsiadlowski, Rappaport \& Pfahl 2002). Dashed
      lines: He-burning star (from Tutukov \& Ferodova 1989). Dashed-dotted lines: the hot C/O white dwarf scenario (Deloye \& Bildsten 
      2003). Dotted lines: the hot He white dwarf scenario (Bildsten 2002, Deloye \& Bildsten 2003).
              }
         \label{fig:modelsJ0929}
   \end{figure}

Using a column density $7.6 \times 10^{20} \leq N_H \leq 1.0 \times 10^{21}$ cm$^{-2}$ (Juett et al. 2003) we estimate a color excess of 
$E(B-V)=0.15 \pm 0.02$ mag and, assuming for the system a distance of 8 kpc, we estimated the expected absolute magnitudes of the 
different possible Roche-lobe filling companion stars and compared them with our photometry of XTE J0929$-$314. 
Following similar work by Jonker et al. (2003) on the AMXP XTE J1751$-$305, 
we considered, for the evolved main-sequence star scenario, the post-period-minimum model for XB 1832$-$330 (${\it M} = 0.026 {\it M_{\odot}}$, 
${\it R} = 0.055 {\it R_{\odot}}$), a binary system with orbital period close to 40 min (Table 1 in Podsiadlowski, Rappaport \& Pfahl 2002) 
and the helium star donor model reported in Table III in Tutukov \& Fedorova (1989; ${\it M} = 0.015 {\it M_{\odot}}$, ${\it R} = 0.046 {\it R_{\odot}}$). 
For the white-dwarf scenario, we used the model of Deloye \& Bildsten (2003) which
estimates for XTE J0929$-$314 either a hot C/O star or a hot He star (${\it M} = 0.011 {\it M_{\odot}}$, ${\it R} = 0.041 {\it R_{\odot}}$ 
and ${\it M} = 0.019 {\it M_{\odot}}$, ${\it R} = 0.050 {\it R_{\odot}}$, 
respectively; Figs. 5, 6 in Deloye \& Bildsten 2003). 
In light of the results obtained above for XTE J1814$-$338 and given the low number of data points, we considered the simple model 
of an irradiated star with a blackbody  spectrum (for details of the modeling see 
Chakrabarty 1998 and Sec. 3.5). As shown in Fig.~\ref{fig:modelsJ0929}, we see that for an irradiating luminosity equal to the X--ray quiescent one 
($L_X \sim 4.5 \times 10^{31}$ erg s$^{-1}$, assuming a distance of 8 kpc; Wijnands et al. 2005) none of the proposed models can 
account for the observed magnitudes of XTE J0929$-$314. To match our photometry, we need an irradiating luminosity higher than about a
factor of 8 ($L_{irr} \sim 3.5 \times 10^{32}$ erg s$^{-1}$). Taking this value as a lower limit for the spin-down luminosity of a 
classical rotating magnetic dipole, we can estimate a lower limit of $1.8 \times 10^8$ G for the neutron star's magnetic field 
(in agreement with the estimate of Monelli et al. 2005 and the upper limit of Wijnands et al. 2005). 

%

\subsection{XTE J1807$-$294}

XTE J1807$-$294 was discovered on  February 21, 2003 revealing a frequency of 190.6 Hz (5.3 ms) by the {\it RXTE} satellite 
during a Galactic plane scan (Markwardt et al. 2003a). A follow-up {\it Chandra} observation allowed to obtain a precise position ($1''$
uncertainty, 90\% c.l.) and to determine an orbital periodicity of 40.1 minutes, making it the shortest orbital period 
of all accreting millisecond X--ray pulsars known to date (Markwardt et al. 2003a). No type I X--ray bursts have been detected, so the
distance of XTE J1807$-$294 is still unknown. \citet{Campana03} observed the source during outburst with the {\it XMM-Newton} 
satellite at a luminosity level of about $2 \times 10^{36}$ erg s$^{-1}$ ($0.5-10$ keV, assuming a distance of 8 kpc) with an X--ray 
featureless spectrum well fitted by an absorbed ($N_H \sim 5 \times 10^{21}$ cm$^{-2}$) blackbody plus hard Comptonization model like 
COMPTT \citep{Titarchuk94}. Kilohertz quasi-periodic oscilllations (kHz QPOs) have been detected in the X--ray flux of XTE J1807$-$294
during outburst with a frequency separation consistent with the 191 Hz pulse frequency \citep{Linares05}. From {\it XMM-Newton}
outburst observations, \citet{Kirsch04} improved the determination of binary orbital parameters and estimated an X--ray mass function
of the system of $f_X(M) = 1.6 \times 10^{-7} M_{\odot}$, which implies (assuming a neutron star mass of 1.4 $M_{\odot}$ and $i =
90^{\circ}$) a minimum companion mass of 0.007 $M_{\odot}$. A re-analysis of these data enabled Falanga et al. (2005) to constrain the
companion mass to $M_c < 0.022 M_{\odot}$, suggesting that is a very low mass (likely white) dwarf.
{\it INTEGRAL} observations of XTE J1807$-$294 in the 0.5$-$200 keV carried out during the 2003 outburst (simultaneously with the 
{\it XMM-Newton} and {\it RXTE} observations) revealed a spectrum consistent with a combination of thermal Comptonization and 
blackbody, as found at lower energy by Campana et al. (2003) but with a higher equivalent temperature \citep{Falanga05b}. 
During quiescence, XTE J1807$-$294 was observed with {\it XMM-Newton} and the source was not detected down to a $3\sigma$ upper limit of
$L_X < 4 \times 10^{31}$ erg s$^{-1}$ (0.5$-$10 keV) assuming a distance of 8 kpc \citep{Campana05}.  
No detection of the counterpart of the system at other wavelengths has been reported so far.

\subsubsection {A search for the optical counterpart during quiescence}

\begin{figure}
\epsfig{file=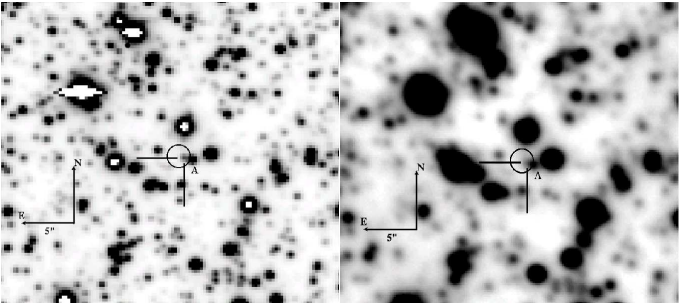, width=9.0cm, height=4.0cm, angle=0}
\vskip 0.0truecm
      \caption{$I-$band ({\it Left}) and $J-$band finding chart for XTE J1807$-$294. The candidate is indicated by solid bars. 
      The black circle represents the $Chandra$ error box. As reported in Sec. 4.2.1, field star ``A'' is probably affecting our optical spectrum, 
      due to its spatial closeness to the target.
              }
         \label{fig:fc1807}
   \end{figure}

Our $VRIJ$ imaging dataset of XTE J1807$-$294 covers, respectively, about 32\%, 32\%, 57\% and 90\% of the 40.1 min orbital period of 
the system (see Table~\ref{tab:log}). An analysis of our $VRI$ images revealed the presence of a source within the precise ($1''$) 
{\it Chandra} error circle (Fig.~\ref{fig:fc1807}) at the following coordinates (J2000): R.A. = $18^h 06^m 59^s.77$, 
Dec = $-29^{\circ} 24' 30''.1$ ($0.3''$ error). As can be seen in Fig.~\ref{fig:fc1807}, this source is barely visible in 
our averaged $J-$band image, obtained under worse seeing conditions, and no solution can be obtained with aperture or PSF-photometry in this band. 
$VRI$ phase resolved PSF-photometry gave no evidence for variability. 
The source appears to be almost constant at a level of $V \sim 22.1$ and $R \sim 21.4$\footnote{When calibrated with 
respect to Landolt standard stars, our $I-$band magnitudes of the candidate counterpart of XTE J1807$-$294 appear to be suspiciously 
brighter than expected from the extrapolated $(V-R)$ color. We found that this behaviour is common to other sources in the field and 
we conclude that our $I-$band photometry is very likely affected by some systematic error and, therefore, we report here only our 
$VR$ results.}. If any variability is present, it should be lower than 0.1 mag.  
Assuming a column density $N_H \sim 5 \times 10^{21}$ (Campana et al. 2003) we estimated a color excess of $E(B-V)=0.79 \pm 0.05$ mag 
and compute the unabsorbed $(V-R) = 0.22 \pm 0.12$ color, typical of an A5-F2 main sequence star.

\begin{figure}
\epsfig{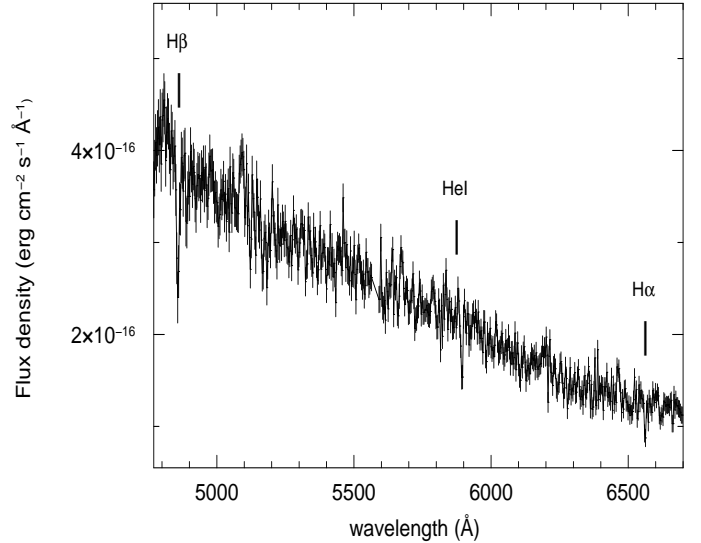}
\vskip 0.0truecm
      \caption{Optical spectrum of candidate counterpart of XTE J1807$-$294 during quiescence (Fig.~\ref{fig:fc1807}). The position of 
      typical optical emission lines of SXRTs are marked.
              }
         \label{fig:spe1807}
   \end{figure}

To further investigate the nature of this source, we obtained on 2006 Sept. 3 an optical spectrum in the wavelength 
range 4000$-$9000 \AA . We extracted the spectrum and found no evidence for Balmer or He emission lines (Fig.~\ref{fig:spe1807}). 
The spectrum is almost featureless, with the exception of $H{\alpha}$ and $H{\beta}$ absorption lines, superposed onto a blue continuum,
suggesting an early-type star (in agreement with the results of our color photometry). 
The fact that the flux density of the spectrum is a factor of $\sim 5$ brighter than our $VR$ photometry (after correcting for 
interstellar reddening) shows that it is probably contaminated by a nearby bright star (star ``A'' in Fig.~\ref{fig:fc1807}). Among the possible scenarios of formation
and evolution of ultracompact X--ray binaries that we have reported in Sec. 4.1, only an evolved main-sequence companion star is supposed
to show hydrogen features in its optical spectrum but they should be detected together with prominent helium lines (Nelemans, Jonker \&
Steeghs 2006). The detection of H absorption features together with the non-detection of He lines in the spectrum of our source and
the absence of variability in the light curve suggests that our candidate is not the quiescent optical counterpart of XTE J1807$-$294 but
more likely an early-type main-sequence star. We then used 3$\sigma$ upper limits of our photometry (reported in
Table ~\ref{tab:photJ1807}) to constrain the possible companion star of XTE J1807$-$294. Assuming a distance of 8 kpc we converted our 
$VRJ$ upper limits into absolute magnitudes (Table ~\ref{tab:photJ1807}).

\begin{table}
\caption{Upper limits ($3\sigma$) on the presence of a source at the X--ray position of XTE J1807$-$294. Values in column three are 
uncorrected for reddening. In column four the dereddening parameters used to correct our optical photometry computed 
assuming $E(B-V)=0.79 \pm 0.05$ mag are reported. The absolute magnitudes are computed assuming a distance of 8 kpc 
(Campana et al. 2005; see Sec. 4.2.1 for details).}
\begin{tabular}{ccccc}
\hline
Filter &${\lambda}_c$&  Limiting magnitude &  A$_{\lambda}$  &   Absolute mag   \\ 
       &   (\AA)     &  	           &     (mag)	     &	                \\ 
\hline
$V$    &  5270       & $> 24.3$            & $2.38 \pm 0.05$   & $> 7.4$        \\
$R$    &  6440       & $> 24.2$            & $1.86 \pm 0.05$   & $> 7.8$        \\
$J$    &  12500      & $> 19.7$            & $0.63 \pm 0.05$   & $> 4.6$        \\
\hline
\end{tabular}
\label{tab:photJ1807}
\end{table}

\begin{figure}
\epsfig{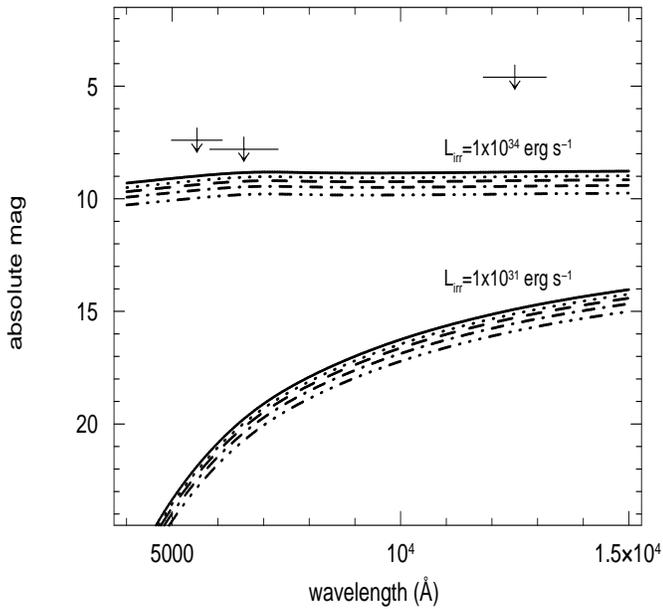}
\vskip 0.0truecm
      \caption{$VRJ$ $3\sigma$ upper limits for the optical counterpart of XTE J1807$-$294 (arrows) compared to the theoretical models for 
      irradiated donor stars in 40-min period ultracompact X--ray binaries (computed for different values of the irradiating luminosity. 
      See Sec. 4.2.1 for details). Symbols used for the different models are the same as Fig.~\ref{fig:modelsJ0929}, with the exception of the
      model of a cold C white dwarf (Deloye \& Bildsten 2003, Falanga et al. 2005), marked with a dashed-dot-dot-dot line.
              }
         \label{fig:modelsJ1807}
   \end{figure}

As done in Sec. 4.1.1 for XTE J0929$-$314, we compared the results of our photometry to a set of spectral energy distributions computed
for the model of an irradiated companion star. All the models we considered for XTE J0929$-$314 (Fig.~\ref{fig:modelsJ0929}) can be
considered valid for XTE J1807$-$294. In addition, we also considered the model of a cold C white dwarf (${\it  M} = 0.011 {\it M_{\odot}}$, 
${\it R} = 0.041 {\it R_{\odot}}$; Deloye \& Bildsten 2003; Falanga et al. 2005). We assumed an irradiating quiescent X--ray luminosity 
of about $1 \times 10^{31}$ erg s$^{-1}$ (consistent with the non-detection of Campana et al. 2005) or a spin-down luminosity of $1 \times 10^{34}$ 
erg s$^{-1}$, as measured for the accretion-powered millisecond X--ray pulsars XTE J1814$-$338 (Sec. 3.5), SAX J1808.4$-$3658 (Campana et al. 2004) 
and IGR J00291+5934 (D'Avanzo et al. 2007).
As shown in Fig.~\ref{fig:modelsJ1807}, our upper limits do not constrain any of the models, even assuming a high irradiating luminosity. 

\subsection{XTE J1751$-$305}

The transient accreting millisecond X--ray pulsar XTE J1751$-$305 was discovered on 2002 April during regular monitoring observations of the Galactic 
bulge region by the {\it RXTE} satellite. The pulsar has a spin frequency of 435.3 Hz (2.3 ms), a 
signature of orbital Doppler modulation on an orbital period of 42.4 minutes; no X--ray eclipses were detected (Markwardt et al. 2002). 
The precise determination of the
orbital parameters enabled  Markwardt et al. (2002) to derive the X--ray mass function $f_X(M) \sim 1.3 \times 10^{-6} M_{\odot}$, which
implies a minimum mass for the companion star of $0.013 - 0.017 M_{\odot}$ depending on the mass of the neutron star. These authors derived
a lower limit for the distance of the source of 7 kpc, by assuming a mass transfer rate from the companion equal to the one measured for SAX
J1808.4$-$3658. {\it XMM-Newton}
observations carried out during the 2002 outburst revealed an X--ray (0.5$-$10 keV) featureless spectrum consistent with a simple absorbed 
blackbody plus power-law model with $N_H = 9.8 \times 10^{21}$ atoms cm$^{-2}$ \citep{Miller03}. \citet{Gierlinski05} presented the
results of an extensive study of all the X--ray data of XTE J1751$-$305 taken with the {\it RXTE} and {\it XMM-Newton} satellites during 
its 2002 outburst. These authors show that the broad X--ray spectrum of XTE J1751$-$305 can be fitted with a model of two soft components
(which originate from a cool accretion disk and a hot-spot on the neutron star surface) and a hard component due to thermal 
Comptonization in a shocked region in the accretion column. During quiescence, XTE J1751$-$305 was observed with {\it Chandra} and was not
detected down to a $0.5-10$ keV luminosity upper limit of $(0.2-2) \times 10^{32}$ erg s$^{-1}$, assuming a distance of 8 kpc (Wijnands et
al. 2005).

\subsubsection {Constraints on the optical counterpart}

The X--ray observations of XTE J1751$-$305 place some strong constraints on the geometry of its companion star. The non-detection of X--ray
eclipses or dips and the assumption that the mass transfer is driven by gravitational radiation (Markwardt et al. 2002) constrain the
orbital inclination to be in the range $30^{\circ}-85^{\circ}$ and the companion mass to be $0.013-0.035 M_{\odot}$. Nelson \& Rappaport
(2003) computed a model for the evolution of a main-sequence, hydrogen-rich companion star of XTE J1751$-$305 that can evolve to the
observed short period of about 42 minutes. According to this model, the current companion star should be a low-mass, low-luminosity 
non-degenerate star with a residual fraction of hydrogen. An alternative scenario is presented by Deloye and Bildsten (2003) and foresees a
hot white dwarf companion with an evolved (He, C/O) composition. The detection (e.g. via optical spectroscopy) of hydrogen would thus be a
key discriminant for determining the validity of these models. Unfortunately, there are no detections of the optical counterpart of 
XTE J1751$-$305 to date, nor in outburst or in quiescence. An observational campaign at optical and NIR bands placed upper limits on the
presence of a star at the precise {\it Chandra} position of XTE J1751$-$305 (Jonker et al. 2003). 

\begin{table}
\caption{Upper limits ($3\sigma$) on the optical counterpart of XTE J1751$-$305; values in column three are 
uncorrected for reddening (from Jonker et al. 2003). In column four are reported the dereddening parameters we used to correct 
for interstellar absorption computed assuming $E(B-V)=1.69 \pm 0.02$ mag The absolute magnitudes are computed assuming a distance 
of 8 kpc (Wijnands et al. 2005; see Sec. 4.3.1 for details).}
\begin{tabular}{ccccc}
\hline
Filter &${\lambda}_c$&  Limiting magnitude &  A$_{\lambda}$  &   Absolute mag   \\ 
       &   (\AA)     &  	           &     (mag)	     &	                \\ 
\hline
$R$    &  6500       & $> 23.1$            & $4.03 \pm 0.02$   & $> 4.6$        \\
$I$    &  8000       & $> 21.6$            & $2.91 \pm 0.02$   & $> 4.2$        \\
$Z$    &  9625       & $> 20.6$            & $2.12 \pm 0.02$   & $> 4.0$	 \\
$J$    &  12500      & $> 19.6$            & $1.35 \pm 0.02$   & $> 3.7$	 \\
\hline
\end{tabular}
\label{tab:photJ1751}
\end{table}

   \begin{figure}
\epsfig{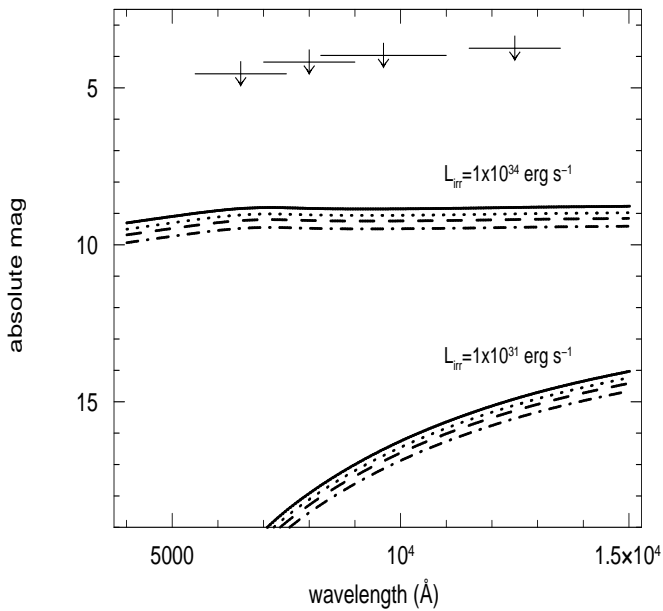}
\vskip 0.0truecm
      \caption{$RIZJ$ $3\sigma$ upper limits on the presence of a source at the X--ray position of XTE J1751$-$305 (arrows) compared 
      to the theoretical models for irradiated donor stars in $\sim$ 40-min period ultracompact X--ray binaries (computed for different 
      values of the irradiating luminosity. See Sec. 4.3.1 for details). Symbols used for the different models are the same of 
      Fig.~\ref{fig:modelsJ0929}.
              }
         \label{fig:modelsJ1751}
   \end{figure}

We used 3$\sigma$ upper limits of the Jonker et al. (2003) photometry (reported in Table ~\ref{tab:photJ1751}) to constrain the possible companion star 
of XTE J1751$-$305 and converted them into absolute magnitudes assuming a source distance of 8 kpc (Table ~\ref{tab:photJ1807}).
Following the procedure discussed in Sec. 4.1.1 for XTE J0929$-$314, we compared the results of Jonker et al. (2003) photometry to different models 
of an irradiated companion star. All the models we have considered for XTE J0929$-$314 (Fig.~\ref{fig:modelsJ0929}) can be
considered valid also for XTE J1751$-$305. We assumed an irradiating quiescent X--ray luminosity of about $1 \times 10^{31}$ erg s$^{-1}$ 
(consistent with the non-detection of Wijnands et al. 2005) and a spin-down luminosity of $1 \times 10^{34}$ erg s$^{-1}$ erg s$^{-1}$.
As shown in Fig.~\ref{fig:modelsJ1751}, these limits are not deep enough to constrain any of the proposed models for the companion star of 
this system, even for a high irradiating luminosity value. 

%

\section{Discussion and Conclusions}

We have presented an extensive study of the optical counterparts of Accreting Millisecond X--ray Pulsars during quiescence. The results of our observational campaigns (also 
presented in Campana et al. 2004 and D'Avanzo et al. 2007), together with data collected from public archives, provided the first comprehensive optical study of the AMXPs 
class during quiescence.
As we discussed in Sec. 1, the eight accreting millisecond X--ray pulsars known to date can be divided into two subclasses 
using the orbital period as a discriminant parameter: compact ($P_{\rm orb} > 1$ hr) and ultracompact ($P_{\rm orb} < 1$ hr) systems. The
compactness of these systems implies very low-mass companion stars as inferred from the measured X--ray mass functions. So, it is reasonable
to foresee a very low optical luminosity during quiescence, when the main contribution to the total optical flux is expected to come 
from the companion star. On the other hand, in such narrow systems, irradiation from the compact object might play a
key role in the observed luminosity, by heating the inner face of the companion star (this energy is then redistributed through the star; \citealt{Podsiadlowski91}). 
We reported the first detection of the quiescent optical/NIR counterparts of XTE J1814$-$338 (Sec. 3.2). Phase resolved photometry of this source revealed a highly 
sinusoidal optical light curve, modulated at the orbital period with a minimum when the neutron star is seen behind the companion, with respect to the observer. 
This behaviour, reminiscent of what observed for SAX J1808.4$-$3658 (Homer et al. 2001; Campana et al. 2004) and IGR J00291+5934 (D'Avanzo et al. 2007), suggests a 
companion star heated by the compact object. This is at variance with the ``classical'' ellipsoidal modulation observed in wider 
LMXTs. The irradiating luminosity required to account for the observed optical flux is in all cases at least one order of magnitude greater than the observed quiescent X--ray 
luminosity. An explanation for this observed ``optical excess'' could be found if we assume that these systems host an active millisecond radio pulsar. According to the 
standard model of formation of millisecond pulsars these neutron stars have been recycled in interacting low-mass X-ray binaries, i.e. have been spun up by accretion after 
substantial decay of their surface magnetic fields. During quiescence (i.e. when the accretion rate onto the neutron star drops by orders of magnitude or even stops) the radio 
pulsar may turn on and start to lose rotational energy in the form of a relativistic particle wind that can irradiate the companion star. 
This wind eventually can be stopped by the pressure of the material flowing from the companion star generating a shock front. 
The presence of such an extended feature may eclipse the light coming from a residual accretion disk (if present during quiescence), providing an explanation for 
the puzzling behaviour of the $V-$band light curve of XTE J1814$-$338 we presented in Fig.~\ref{fig:lc} and discussed in Sec.~3.5. 

As for the compact AMXP systems, the detection of the observed optical flux during quiescence of the ultracompact system XTE J0929$-$314 can be accounted 
for only by invoking an irradiating luminosity about one order of magnitude greater than the quiescent X--ray luminosity (Fig. 8).
Taking our estimates of the irradiating luminosity as a lower limit for the spin-down luminosity of a classical rotating magnetic dipole, we could estimate the 
neutron star's magnetic field for both systems. The derived limits are in agreement with the standard scenario of weakly magnetized
($10^7-10^8$ G) neutron stars to be the progenitors of millisecond radio pulsars (Sec. 3.5 and 4.1.1). 
We note that no radio pulsations were detected from XTE J0929$-$314 during quiescence. Beaming factor and intrinsic low radio luminosity were 
suggested as the most likely explanations for this non-detection \citep{Iacolina09}.
XTE J1751$-$305 and XTE J1807$-$294 lie both on the Galactic plane and are therefore heavily absorbed (we estimated $A_V \sim 5.2$ mag and $A_V \sim 2.4$ mag, respectively). 
Even assuming a high spin-down luminosity ($L_{irr} = 1 \times 10^{34}$ erg s$^{-1}$), the upper limits of the photometry are consistent with a non detection of the quiescent 
optical counterpart. 

As we discussed in Sec. 4, there are two main basic and competing scenarios for the formation of ultracompact binary systems. The first involves an evolved
main-sequence or a He-burning companion star (both with a residual presence of hydrogen) while the second predicts for the presence of a He or C/O white dwarf. We note
that according to both scenarios the companion star of these ultracompact systems should be a very low-mass ($M_c \sim 0.01 \, M_{\odot}$) low-luminosity star.  
As shown in Figures 7, 10 and 11, with our data we cannot constrain any of the evolutionary models for ultracompact binary systems. However, as we reported in Sec. 4, 
the detection of hydrogen from one of these systems would be a key discriminant in determining the validity of the first scenario (which involves an evolved main-sequence or a 
He-burning star) with respect to the second one (white dwarf companion). Optical spectroscopy of XTE J0929$-$314 carried out during outburst did not reveal any clear 
evidence of hydrogen or helium lines (Nelemans, Jonker and Steeghs 2006). In addition, we note that, at variance with what is observed for all the compact systems, none of 
the ultracompact AMXPs showed Type-I X--ray bursts. These events originate from explosive nuclear ignition on the surface of neutron stars. At a critical point, 
degenerate hydrogen and/or helium burning ignites explosively, suddenly heating up the entire surface, enough to emit strong X-rays \citep{Maraschi77,Woosley76}. 
The absence of Type-I bursts among ultracompact AMXPs finds a natural explanation if we assume that there is no hydrogen and/or He left in the donor star, as expected in 
the case of a C/O white dwarf companion star, suggesting that the other evolutionary channel for these systens is not populated. 

With quiescent unabsorbed magnitude in the range $R \sim 22-23$, compact AMXPs are well within the 
spectroscopic capabilities of telescopes of the 8-meter class. An estimate of the radial velocity of the companion star can lead to a determination of the optical mass 
function and, consequently, to place constraints on the mass of the two stars. Firm upper limits on the neutron star mass could be fundamental to discriminate between 
different equations of state. Phase-resolved spectroscopy of these sources would be the natural continuation of the work presented in this paper.

\begin{acknowledgements}
We thank the anonymous referee for her/his useful comments and suggestions.
PDA and SC acknowledge the Italian Space Agency for financial support through the project ASI I/R/023/05. PDA wants to thank Prof. A. Treves for useful suggestion and discussion.
PDA acknowledge H. Sana, S. Hubrig and L. Faunez for useful help given during observations at the ESO-VLT.
\end{acknowledgements}

\bibliographystyle{aa}
\bibliography{bib_amxp_s}

\end{document}